# Visualizing Band Offsets and Edge States in Bilayer-Monolayer Transition Metal Dichalcogenides Lateral Heterojunction

*Chendong Zhang[1], Yuxuan Chen[1], Jing-Kai Huang[1,5], Xianxin Wu[2,4], Lain-Jong Li[5], Wang Yao[2], Jerry Tersoff[3], and Chih-Kang Shih[1*]*

[1]*Department of Physics, University of Texas at Austin, Austin, TX 78712, USA*
[2] *Department of Physics and Center of Theoretical and Computational Physics, University of Hong Kong, Hong Kong, China*
[3]*IBM Research Division, T. J. Watson Research Center, Yorktown Heights, NY 10598, USA*
[4]*Institute of Physics, Chinese Academy of Sciences, Beijing 100190, China*
[5]*Physical Sciences and Engineering Division, King Abdullah University of Science and Technology, Thuwal, 23955-6900, Kingdom of Saudi Arabia*

*\*Corresponding author E-mail: shih@physics.utexas.edu*

**Semiconductor heterostructures are fundamental building blocks for many important device applications. The emergence of two-dimensional semiconductors opens up a new realm for creating heterostructures. As the bandgaps of transition metal dichalcogenides thin films have sensitive layer dependence, it is natural to create lateral heterojunctions using the same materials with different thicknesses. Using scanning tunneling microscopy and spectroscopy, here we show the real space image of electronic structures across the bilayer-monolayer interface in $MoSe_2$ and $WSe_2$. Most bilayer-monolayer heterojunctions are found to have a zigzag-orientated interface, and the band alignment of such atomically sharp heterojunctions is of type-I with a well-defined interface mode which acts as a narrower-gap quantum wire. The ability to utilize such commonly existing thickness terrace as lateral heterojunctions is a crucial addition to the tool set for device applications based on atomically thin transition metal dichalcogenides, with the advantage of easy and flexible implementation.**

A heterojunction is an interface between two different semiconductors. The difference in the electronic structures of the two materials results in potential discontinuities at the interface for electrons (conduction band offset) and holes (valence band offset). The introduction of heterostructures[1] has enabled new types of electronic, photonic, transforming semiconductor technology. The recent emergence of two-dimensional (2D) semiconductors creates exciting new opportunities to push semiconductor heterostructures toward a new frontier[2-7].

Vertically stacked van der Waals heterostructures have been quickly recognized as a powerful platform to create atomically thin heterostructures with great design flexibility[2,8-10]. Indeed, van der Waals heterostructures have been realized using different combinations of 2D materials, including transition metal dichalcogenides (TMDs), graphene and boron nitride [11-15]. Interesting properties, such as the observation of interface excitons[16] and the determination of band alignments have been demonstrated recently in TMDs vertical heterostructures[17]. Less attention has gone to the possibility of creating "lateral heterojunctions", where the junction is now a line interface between two 2D materials, reducing the dimensionality of heterostructures even further. Using direct chemical vapor deposition (CVD) growth, lateral junctions between different TMDs have been recently demonstrated[18-22]. However, theoretical calculations suggest that the heterostructures formed between the four most common TMD compounds (M= Mo, W; X = S, Se) all have type-II band alignment[23,24], and the equally (if not more) important type-I heterojunctions (HJs) are still missing. How to create a lateral type-I heterojunction with an atomically sharp and straight interface, and characterize its band profile at the atomic scale, remain as significant challenges to overcome in order to advance this new frontier based on 2D semiconductors.

Previous investigations have shown that the bandgaps of TMD films greatly depend on the number of layers[25,26]. So one may naturally ask: Can we create a well-defined lateral heterojunction between regions of different thicknesses, e.g. a bilayer – monolayer interface? If so, is such a HJ type-I or type-II? Moreover, as the junction interface is also a step edge, how does this step edge influence the electronic structures of the heterojunction?

Here we study this new type of lateral heterostructures between bilayer (BL) and monolayer (ML) TMDs using low temperature scanning tunneling microcopy and spectroscopy (STM/S), We first show that atomically sharp and smooth interface with the desired type-I band alignment can indeed be formed. Moreover, we discover the presence of interface states with a narrow gap that act as interface quantum wires. Because the junction is formed between the direct gap monolayer TMD and the indirect gap bilayer TMD, the band edge carriers on the two sides of the junction are from different momentum space regions. This feature, coupled with the ability to control the interface through the edge states, offers the opportunity for novel device concepts in reduced dimensions.

**Results**

**WSe$_2$ bilayer – monolayer heterojunctions.** The BL-ML lateral heterojunctions are formed naturally when a second layer of TMD is grown on top of the ML TMD. We achieve direct growth of such a HJ either using CVD[27] or molecular beam epitaxy (MBE)[26,28]. Shown in Fig. 1a is an example of a WSe$_2$ BL-ML heterostructure grown on a highly-oriented-pyrolytic-graphite (HOPG) substrate using CVD. In this scanning tunneling microscopy (STM) image, three distinct regions – graphite, ML-WSe$_2$ and BL-WSe$_2$ – are separated by two atomically sharp line interfaces, corresponding to the ML and BL step edges (respectively separating graphite from ML WSe$_2$, and ML from BL WSe$_2$). An atomically resolved STM image (inset) taken on the BL region allows us to determine the orientation of these two interfaces to be along the zig-zag directions. Then scanning tunneling spectra are acquired to reveal the real space band profile across the interface.

Figure 2a is a close-up STM image showing the spatial locations where scanning tunneling spectroscopy given in Fig. 2b and 2c is carried out. Shown in Fig. 2b is a color rendering of the band mapping result, and selected individual spectra are shown in Fig. 2c. Note that d$I$/d$V$ spectra are displayed on a logarithmic scale. The numbers on the spectra refer to their position in the complete set, in which spectra are acquired every 2 nm from the BL into the ML region, with spectrum # 14 acquired at the interface (see Fig. 2a).

The most striking feature in Fig. 2b is the apparent band bending for both conduction band and valence band in the BL WSe$_2$ near the interface. The magnitude of the band bending $\Delta_{bend}$ is about 0.15 eV over about 10 nm toward the interface (depletion length). One can also see an abrupt change in the electronic structure right at the BL-ML interface (see also spectrum #13 and #14 in Fig. 2c). The spectra in the ML region, in contrast, exhibit little band bending.

In the BL region, the d$I$/d$V$ spectrum shows a prominent peak in the valence band (labeled with black arrow in spectrum #1 of Fig. 2c). This corresponds to the higher energy branch at the Γ point split due to the interlayer coupling. This feature has been discussed in detail recently[29]. The lower state is not visible in the limited energy window here. Far away from the interface, the valence band maximum (VBM) of BL region is located at -1.10 ± 0.07 eV while the conduction band minimum (CBM) is determined to be 0.71 ± 0.07 eV (from spectrum #1-#6), corresponding to a quasiparticle band gap of 1.81 ± 0.10 eV. From #6 to #12, the spectral line shape remains the same but the location of the VBM and CBM both move upward, corresponding to the band bending (as indicated by the arrows in #6 and #12).

Right at the BL-ML interface, new spectral features emerge (#13 and #14 in Fig. 2c). Two prominent peaks around 0.4 eV and 0.8 eV (marked by the red arrows) appear in the conduction band. The new spectral features in the valence band can be observed more vividly on spectrum

#14 (marked by the green arrows) but merge with the bulk valence states in #13. Interestingly, as revealed by the STS, the interface states here behave effectively like a narrow gap semiconductor with a gap value of 0.8 eV (with VBM and CBM located roughly at -0.6 eV to 0.2 eV respectively). Our STS map ( Fig. 2b) suggests that the interface states have type-I band alignment with 2D bulk states of both the BL and ML therefore can play the role of a quantum wire for both electrons and holes. Localized interface states are actually a common feature of semiconductor interfaces having broken bonds, such as surfaces states at crystal-vacuum interfaces. In our case, broken bonds occur at the termination of the upper layer of bilayer TMD. Thus we may expect the quantum wire states to be centered on the atoms terminating the edge. However, the details of the step structure are not yet fully known.

Moving into the ML $WSe_2$ region, one quickly observes the bulk ML $WSe_2$ electronic structure without visible band bending within the spatial resolution (2 nm) of the STS. As one can see from spectrum # 15 to # 30 the electronic structure remains constant. It is important to recognize that in the ML $WSe_2$ region, it is difficult to directly observe the actual VBM location (at K points in Brillouin zone) using conventional constant $Z$ tunneling spectroscopy as in Fig. 2b and 2c, due to the extremely short tunneling decay length of K-points states[29]. The peak in the valence band corresponds to the energy location of the $\Gamma$ point while the VBM at K points (not visible in the constant Z spectroscopy) is approximately located at 0.65 eV above. This point will be discussed further below.

As in any heterojunction, the energy alignment of the band edges (including the interface quantum wire) right at the interface is an inherent property, while the band bending depends on the electrostatics (for example the substrate). In the bulk ML and BL regions, the intimate contact with the graphite determines the location of $E_F$ in the gap. Near the edge, band bending occurs in order to keep the $E_F$ in the band gap of the zig-zag quantum wire. The screening by the substrate graphite sets the length scale of the band bending at about 10 nm.

**MoSe$_2$ bilayer – monolayer heterojunctions.** We have also studied the BL-ML heterostructure in MBE-grown MoSe$_2$ on HOPG, where we find the same general behavior as for WSe$_2$, as illustrated in Fig. 3. The spatially resolved STS is acquired with a finer step of 0.8 nm (labeled in Fig. 3a). The color rendering of band mapping based on the STS is shown in Fig. 3b. The upward band bending is observed for both conduction and valence bands. However, the "apparent" band bending from the valence band side is much stronger, effectively reducing the band gap starting about 3-4 nm away from the edge. On detailed inspection, this is due to the overlap of valence band states of the bulk with those of the interface quantum wire states in the STS signal. One can see this more clearly from the individual STS spectra shown in Fig. 3c. Spectrum #10 still resembles the bulk BL band structure. On the other hand, in spectrum #11 the interface states above the bulk valance band edge start emerging and eventually split off (labeled by the green arrows in spectrum #15). This behavior is similar to that in WSe$_2$ discussed above except here the magnitude of the energy split off from the VBM of the bilayer is larger. The conduction band interface state is only observed here right at the interface (#15, labeled by the red vertical arrow). This might be due to a smaller energy split off from the CBM of the bilayer making it only observable very near the edge. The edge here has a narrower gap of 0.4 eV, which is also smaller than the one in WSe$_2$. Moreover, with finer spatial resolution, we can observe a small band bending in ML region with a very short depletion length (around 1 nm only).

We summarize the general behavior of lateral heterojunctions formed from BL and ML WSe$_2$ or MoSe$_2$ with the schematic diagram (Fig. 4a) and Table 1. Here the amount of the energy band bending is labeled as $\Delta_{Bend}$, the gap of the interface mode as $\Delta_{Intf}$, the valence band offset as VBO and the conduction band offset as CBO. The numerical values for these quantities are shown in Table 1 for WSe$_2$ and MoSe$_2$, respectively. As mentioned above, in the ML region, constant height tunneling spectroscopy lacks the sensitivity to reveal the location of the VBM at the K-point. On the other hand, as we reported recently[29], constant-current spectroscopy can overcome this difficulty and resolve the states at the K point in the valence band (labeled as K$_V$). As discussed extensively in ref. 29, in such constant-current spectroscopy, the individual thresholds at different critical points appear as peaks in differential conductivity $(\partial I/\partial V)_I$ as well as dips in the derivative of the tip-to-sample position as a function of the voltage $(\partial Z/\partial V)_I$. Shown in Fig. 4b is the case for WSe$_2$. The actual VBM is located at the mid-point of the transition from the TMD to the graphite states, which is around 0.65 eV above the Γ point, as labeled in Fig. 2b. For MoSe2, this value is 0.40 ± 0.04 eV. With this information, we are able to determine that the CBO and VBO of BL-ML WSe$_2$ HJs are 0.15 ± 0.10 eV and 0.12 ± 0.10 eV, respectively as shown in Table 1. Similarly, for BL-ML MoSe$_2$, they are found to be 0.08 ± 0.10 eV and 0.43 ± 0.10 eV, albeit in this case, the determined CBO is smaller than the experimental uncertainty of 0.1 eV.

**Discussion**

We first discuss the implications for electron transport across or along the interface that may raises many interesting issues. It is well known for semiconductor heterojunctions that there can be strong reflection at the interface, depending on how dissimilar the two material are. The same is true for reduced-dimensionality interfaces, such as heterojunctions between monolayer and bilayer graphene[30,31]. Considering the significant changes of their electronic structures (especially band edges locating at different **k** points in Brillouin zone), we may expect similar strong reflection at the interfaces of these BL-ML TMDs heterojunctions. Transport along the interface can also be subject to scattering by imperfections. For example, the edge of the upper layer is in general not perfectly straight, so transport along the quantum wire will experience some disruption at kinks and other imperfections in the edge structure. This could lead to hopping conductivity with poor mobility along the wire. However, the non-ideal transport need not be an obstacle to important applications. In many conventional applications, the primary role of the heterojunction is to provide confinement, or to control barriers; and transport along or across the interface is secondary or even irrelevant. More important is that these interfaces provide a rich combination of barriers and localized states, offering novel opportunities for device engineering. Moreover we anticipate that the localized quantum wire states could be easily doped, since atoms tend to diffuse and bind at step edges.

In conclusion, our STM/STS studies reveal a new class of lateral heterojunction in atomically thin $WSe_2$ and $MoSe_2$, formed from naturally existing ML-BL thickness terrace with zigzag orientation. Because of the thickness dependence of the band gap and band edges, the ML and BL bulk bands align to form a type-I heterojunction, as clearly seen in our STS maps. The interface states in both systems form a narrow gap quantum wire. This interface quantum wire has a gap value of 0.8 eV in $WSe_2$ and a smaller gap value of 0.4 eV in $MoSe_2$. The band alignment of the quantum wire states is also type-I with respect to both the BL and ML sides of the junctions. We expect that the unique properties of this novel class of heterojunctions will create new possibilities for device applications based on 2D TMDs.

## Methods

**Growth of 2D TMDs samples.** The preparation of WSe$_2$ crystal flakes by the vapour-phase reaction has been reported before[27]. In brief, high purity metal trioxides WO$_3$ was placed in a ceramic boat in the center of a furnace while graphite substrate was placed in the downstream side of the furnace, adjacent to the ceramic boat. Selenium powder was heated by a heating tape and carried by Ar or Ar/H$_2$ gas to the furnace heating center. The temperature of furnace was gradually raised from room temperature to the desired temperature, and cooled down naturally after the reaction had occurred. MoSe$_2$ was grown on freshly cleaved HOPG substrate using MBE in an ultra-high-vacuum (UHV) chamber which has a base pressure of $5 \times 10^{-11}$ Torr. High purity Mo (99.95%) and Se (99.999%) were evaporated from a home-built e-beam evaporator and an effusion cell, respectively, with a ratio of 1:30. The graphite substrate was kept at 550 ºC, and the growth rate was about 0.3 layer/hour. The sample was annealed in a Se flux at 600 ºC for 30 min after growth. Before STM studies, the CVD samples are cleaned in the UHV chamber (base pressure is lower than $6 \times 10^{-11}$ Torr) by annealing the sample at 300 ºC for 6 hours. The MBE samples are transferred *in situ* between the growth chamber and the STM chamber under UHV environment.

**Scanning tunneling microscopy and spectroscopy.** All STM investigations reported here were acquired at 77 K in ultra-high-vacuum (base pressure is lower than $6 \times 10^{-11}$ Torr). Electrochemically etched tungsten tips were cleaned in-situ with electron beam bombardment. The tunneling bias is applied to the sample. The conductance spectra were taken by using a lock-in amplifier with a modulation voltage of 10 mV and at a frequency of 924 Hz.

**Acknowledgements**

This research was supported with grants from the Welch Foundation (F-1672), and the US National Science Foundation (DMR-1306878, EFMA-1542747). W.Y. thanks the support from RGC of HKSAR (HKU9/CRF/13G) and HKU R.O.P. L.J.L. thanks the support from KAUST (Saudi Arabia), MOST and TCECM, Academia Sinica (Taiwan) and AOARD-134137 (USA). C.K.S is partially supported by the NT 3.0 Program, Ministry of Education.

**Author Contributions**

C.K.S., C.D.Z. and J.T. conceived the project. C.D.Z. carried out the STM/S measurement. Y.X.C. carried out MBE growth of $MoSe_2$, J.K.H. and L.J.L. grew $WSe_2$ by CVD. X.X.W. provided computational support for understanding the edge states. C.D.Z., C.K.S., W.Y., and J.T. analyzed the data, interpret the data and wrote the paper.

**Additional information**

**Competing financial interests**

The authors declare no competing financial interests.

**Figure Legends**

**Figure 1 | STM images and schematic models of BL-ML TMD lateral heterojunctions.** (**a**) STM image for a naturally formed BL-ML WSe$_2$ HJ grown on HOPG by chemical vapour deposition. The bilayer and monolayer regions are labelled as shown. (**b**) Atomic resolution image taken on bilayer region where a dash square is labelled in **a** (not to scale). The sample biases and tunnelling currents used are (**a**) 3V, 8 pA, (**b**) -1.0V, 10 pA. The arrows in (**a**) and (**b**) indicate the zig-zag orientation of the bilayer-monolayer interface. (**c,d**) Schematic models of the BL-ML heterojunction viewing from top and side, respectively. The green and cyan represent the metal atoms in second and first layer respectively, while the red and yellow corresponding for chalcogen atoms. Scale bar, (**a**) 100 nm and (**b**) 1 nm.

**Figure 2 | Scanning tunnelling spectroscopy investigation of band profile across the BL-ML WSe$_2$ heterojunction.** d$I$/d$V$ spectra were taken along the path shown in (**a**). The spectra numbers are labeled (counting from left to the right in the path line). The total length is roughly 73 nm with a step size of 2 nm. Spectrum #14 was taken right at the interface. (**b**) Color coded rendering of the real space imaging of band profile plotted in terms of Log(d$I$/d$V$). All spectra in this paper are displayed with arbitrary units (a.u.), except for Fig. 4(c). The energy locations of Γ and K$_V$ (actual VBM) points are labelled by black arrows. (**c**) A selective subset of Log(d$I$/d$V$) spectra. In spectrum #1, the upper state of Γ splitting, which results from interlayer coupling, is labelled with a black dashed arrow. The interface states are marked in spectrum #13 and #14 with red arrows for conduction band side and green arrows for valence band side. A black arrow in monolayer region represents for the energy location of CBM. Scale bar, (**a**) 20 nm.

**Figure 3 | Scanning tunnelling spectroscopy investigation of band profile across the BL-ML MoSe$_2$ heterojunction.** Similar with Figure 2, (**a**) is a close-up STM image showing the path line which spectroscopy was taken along. The total length is about 18 nm with a step size of 0.8 nm. The spectra numbers in (**b**) and (**c**) are counted from left to right in the path, while spectrum #15 was taken right at the BL-ML interface. (**b**) Color coded rendering of the real space imaging of band profile plotted in terms of Log(d$I$/d$V$). (**c**) Selective subset of Log(d$I$/d$V$) spectra. In spectrum #15, the interface states in valence band and conduction band are marked with green and red arrows, respectively. Scale bar, (**a**) 5 nm.

**Figure 4 | Schematic diagram of band alignments and constant Z spectroscopy for valence band of ML WSe$_2$.** (**a**) Schematic diagram showing the generic band alignment in BL-ML TMDs HJs. The magnitude of the band bending is labelled as $\Delta_{Bend}$, and the gap of interface states is $\Delta_{Intf}$. (**b,c**) Two modes of constant $Z$ spectroscopy – $(\partial I/\partial V)_I$ and $(\partial Z/\partial V)_I$ for valence band of monolayer WSe$_2$. The actual valence band maximum at K point is labeled as K$_V$ which is about 0.65 eV above Γ point. See reference 29 for details.

| BL-ML | $\Delta_{Bend}$ (BL) | CBO | VBO | $\Delta_{Intf}$ |
|---|---|---|---|---|
| WSe$_2$ | 0.15 ± 0.05 eV | 0.15 ± 0.10 eV | 0.12 ± 0.10 eV | 0.8 ± 0.10 eV |
| MoSe$_2$ | | 0.08 ± 0.10 eV | 0.43 ± 0.10 eV | 0.4 ± 0.10 eV |

**Table 1 | Band alignments in bilayer-monolayer heterojunctions for WSe$_2$ and MoSe$_2$.** The standard deviations shown here are based on statistics of multiple measurements (more than 50 times).

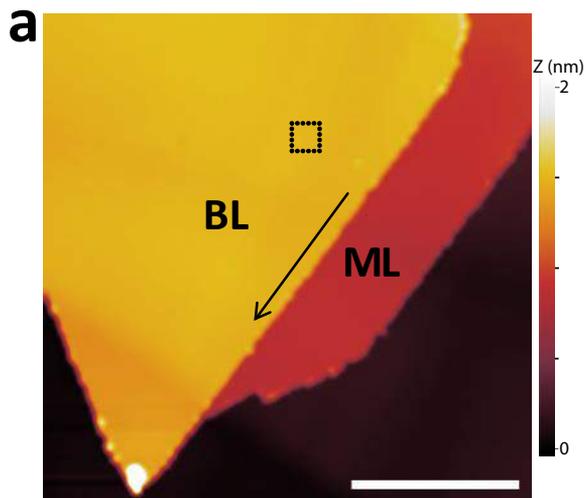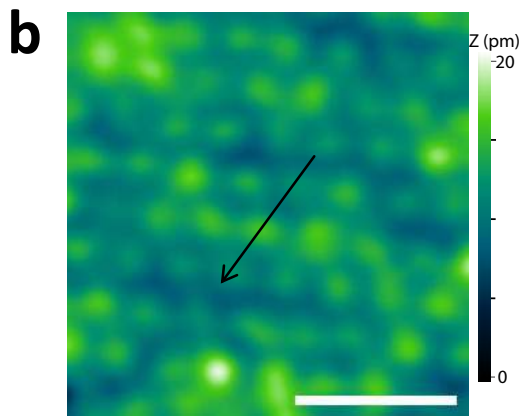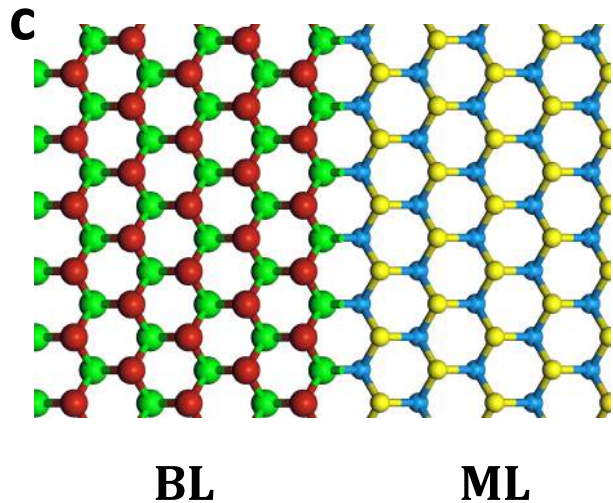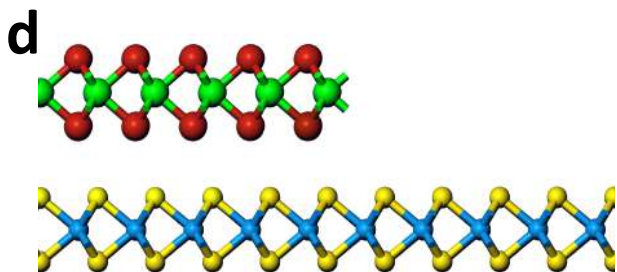

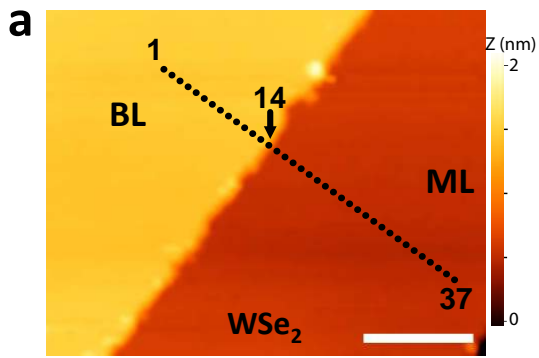
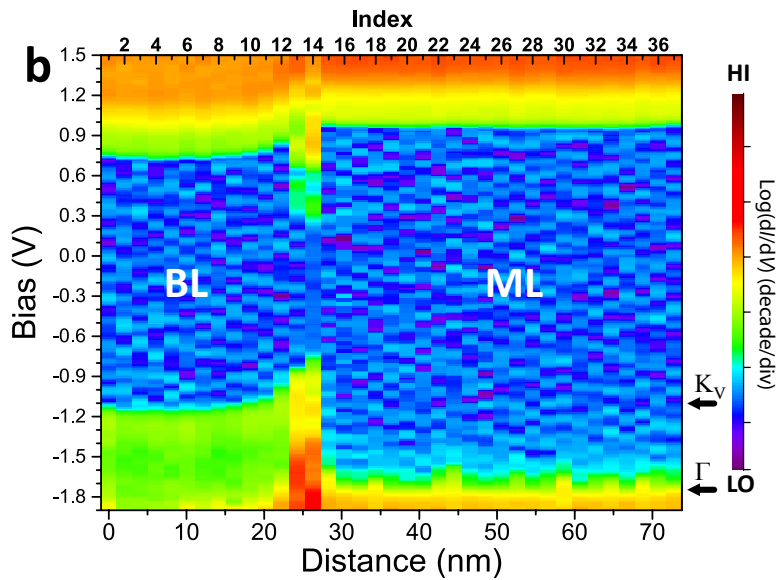
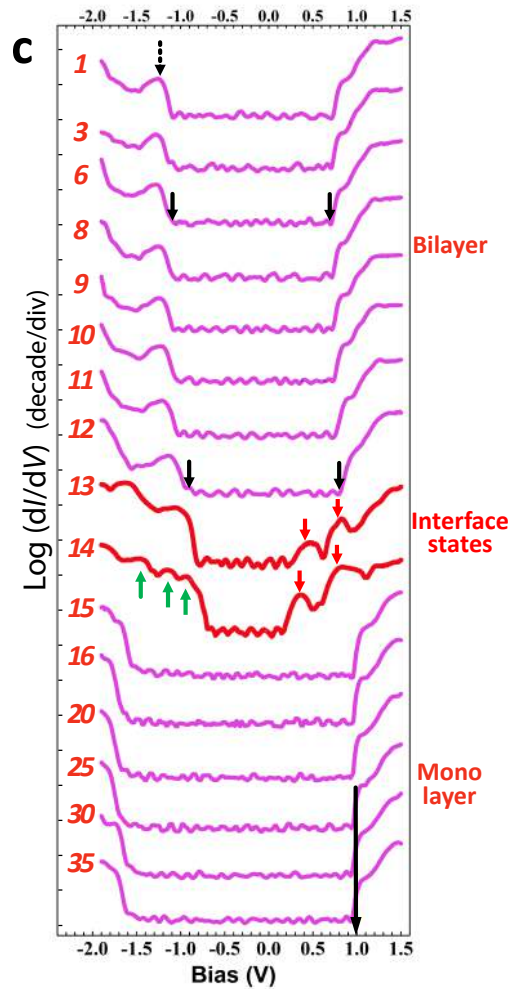

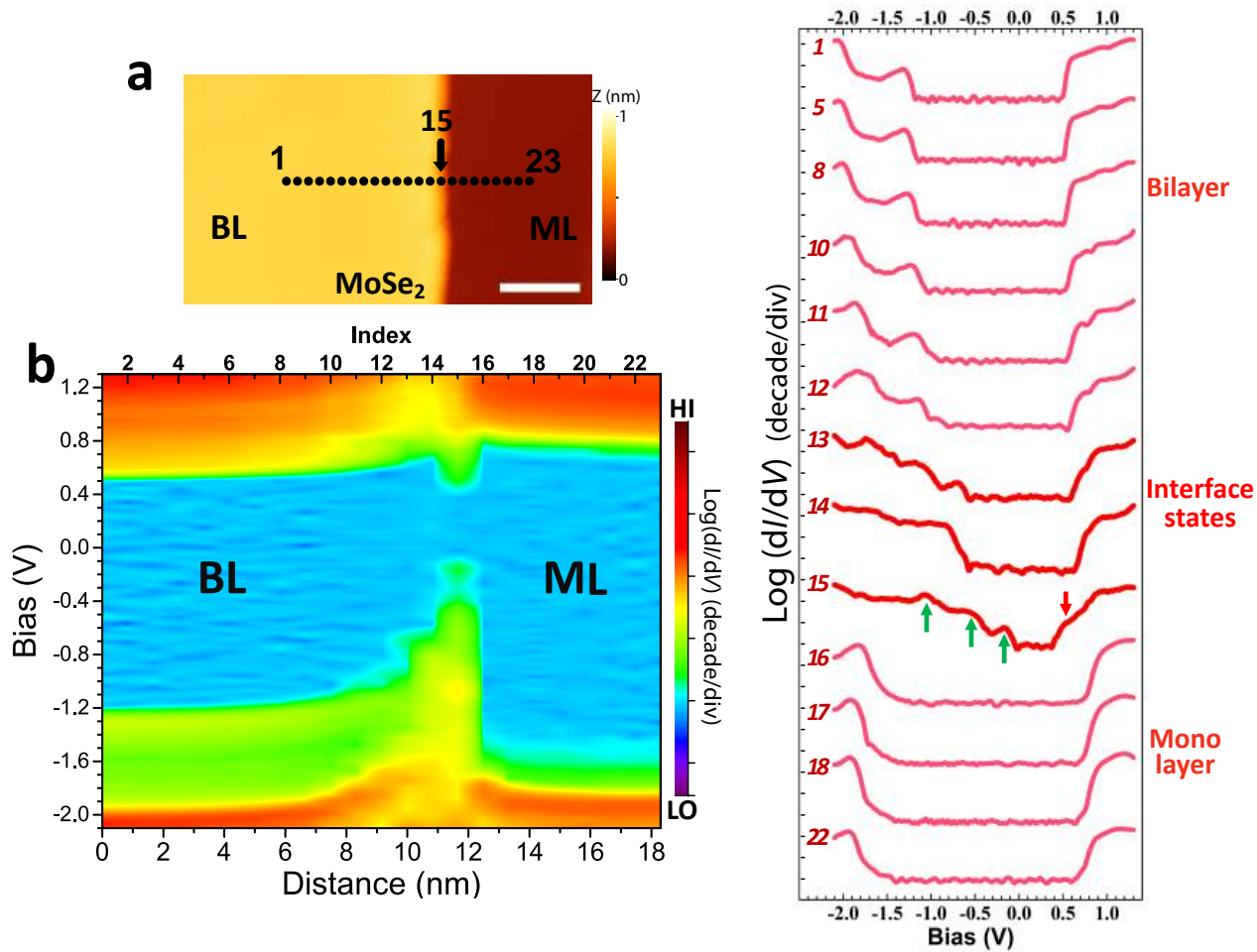

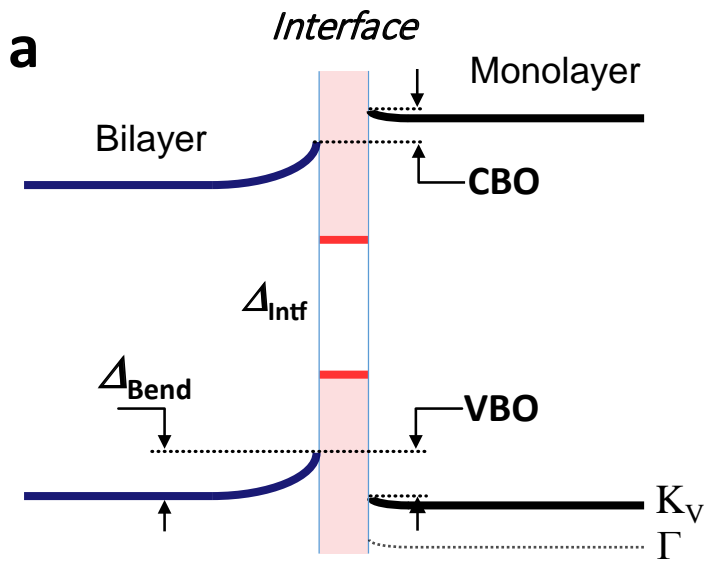
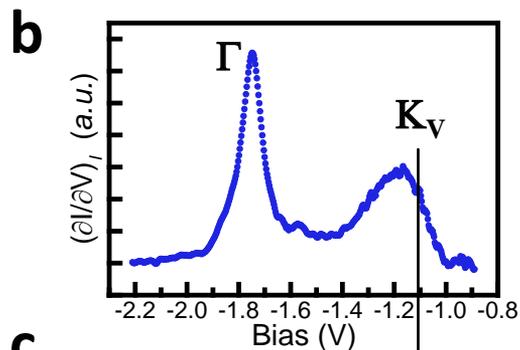
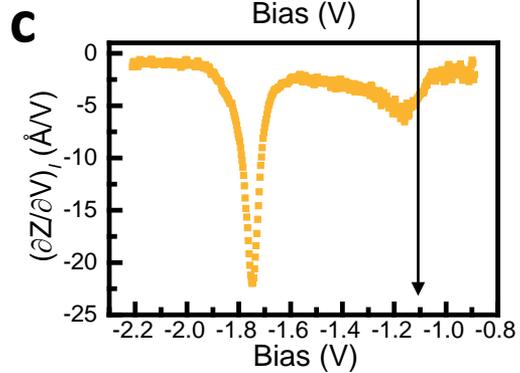